\documentclass[a4paper,11pt]{article}
\pdfoutput=1
\usepackage{jinstpub}
\title{\boldmath Neutron imaging and tomography with \textsc{mcp}s}
\author[a,1]{S. Duarte Pinto\note{Corresponding author.},}
\author[a]{R.~Ortega,}
\author[a]{S.~Ritzau,}
\author[a]{D.~Pasquale,}
\author[a]{B.~Laprade,}
\author[a]{S.~Mrotek,}
\author[a]{S.~Gardell,}
\author[b]{Z.~Zhou,}
\author[b]{J.~Plomp,}
\author[b]{L.~van Eijck,}
\author[c]{H.~Bilheux,}
\author[c]{and I.~Dhiman}
\affiliation[a]{\textsc{Photonis} Technologies S.A.S.,\\Avenue de Pythagore, 33700 M\'erignac, France}
\affiliation[b]{Delft University of Technology,\\ Mekelweg 15, 2629 JB Delft, the Netherlands}
\affiliation[c]{Oak Ridge National Laboratory,\\ 1 Bethel Valley Road, Oak Ridge, TN 37830, USA.}
\emailAdd{S.DuartePinto@photonis.com}

\abstract{A neutron imaging detector based on neutron-sensitive microchannel plates (\textsc{mcp}s) was constructed and tested at beamlines of thermal and cold neutrons.
The \textsc{mcp}s are made of a glass mixture containing $^{10}$B and natural Gd, which makes the bulk of the \textsc{mcp} an efficient neutron converter.
Contrary to the neutron-sensitive scintillator screens normally used in neutron	imaging, spatial resolution is not traded off with detection efficiency.
While the best neutron imaging scintillators have a detection efficiency around a percent, a detection efficiency of around 50\% for thermal neutrons and 70\% for cold neutrons has been demonstrated with these \textsc{mcp}s earlier.

Our tests show a performance similar to conventional neutron imaging detectors, apart from the orders of magnitude better sensitivity.
We demonstrate a spatial resolution better than 150 \textmu m.
The sensitivity of this detector allows fast tomography and neutron video recording, and will make smaller reactor sites and even portable sources suitable for neutron imaging.}
\arxivnumber{1710.02614}
\proceeding{\\19$^{\text{th}}$ International Workshop on Radiation Imaging Detectors (iWoRiD)\\ 2--6 July 2017\\ Krakow, Poland}

\begin{document}
\noindent\includegraphics[width=35mm]{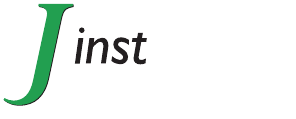}\hfill
\maketitle
\section{Introduction}
Detectors for thermal neutron imaging need to combine a good spatial resolution with a field of view sufficiently wide for the sample under study.
The most conventional technology combining these requirements is a thin neutron-sensitive scintillator, read by a highly sensitive scientific \textsc{cmos} or \textsc{ccd} camera.
The scintillating material is usually ZnS:Cu, emitting around 531 nm, near the optimum of s\textsc{cmos} sensitivity.
This scintillator is used in powder form, also known as the green phosphor P31, mixed with $^6$LiF powder.
The $^6$Li (highly enriched) here acts as the neutron converter through the fission reaction $\textrm{n} (^6Li,\alpha) ^3\textrm{H}$.
Both fission fragments have a range of few tens of microns in the compound.
The resolution is limited by the thickness of the $^6$LiF/ZnS:Cu layer; with both constituents in powder form the layer has a white appearance, and light scattering between grains limits the resolution to approximately the thickness of the layer.

This way of constructing an imaging detector has many benefits.
It offers great flexibility: $^6$LiF/ZnS:Cu is inexpensive, and can be applied by brushing or spraying on aluminum substrates of any size, which are transparent to neutrons.
If the thickness of the layer is well controlled, a resolution of better than 50 microns can be attained.
All elements in the mixture have a low atomic number, which makes it rather insensitive to the gamma background found in every neutron beam environment.

A difficulty with this technique, however, is the limited sensitivity, and the fact that sensitivity is necessarily traded off with spatial resolution.
Scintillators with thickness from fifty to a few hundred microns typically stop only a few percent of thermal neutrons.
Moreover, ZnS:Cu has a long afterglow\footnote{ZnS:Cu is also known as the \emph{glow-in-the-dark} phosphorescent in many toys and cosmetics.}, complicating tomography or other situations where multiple consecutive images are taken.
An alternative neutron scintillator sometimes used, Gadox (Gd$_2$O$_2$S), does not have this afterglow, but due to the high atomic number of gadolinium ($\textrm{Z}=64$) it is also quite efficient as a gamma scintillator, contributing to the background.

In this work we present a neutron imager based on a neutron-sensitive microchannel plate (\textsc{mcp}) coupled to a fast phosphor.
The physical mechanism by which neutron capture in the \textsc{mcp} leads to a scintillation image on the screen is quite different from the case with neutron-sensitive scintillators.
But since in both cases the result is an image on a scintillating screen, our MCP imager can be placed in the position of a neutron scintillator, and be read with the same mirror/lens/camera setup already in use.
This is indeed how we have tested our prototype.
The benefit of this system compared to neutron scintillators is primarily a high spatial resolution that is combined with a good detection efficiency; efficiency and resolution are not traded off.

The possibility to make \textsc{mcp}s of a modified glass composition to make them neutron sensitive was proposed and demonstrated by Fraser in 1990 \cite{Fraser}.
This concept was further studied and optimized by Tremsin et al., who eventually demonstrated a detection efficiency of 50\% for thermal neutrons and 70\% for cold neutrons \cite{Tremsin_efficiency}, and a spatial resolution better than 15 \textmu m \cite{Tremsin_resolution}.
These same optimized \textsc{mcp}s are the basis of the neutron imager presented here.

\section{An \textsc{mcp}-based neutron imager}
 \begin{figure}[tb]
	\centering\includegraphics[width=.68\textwidth]{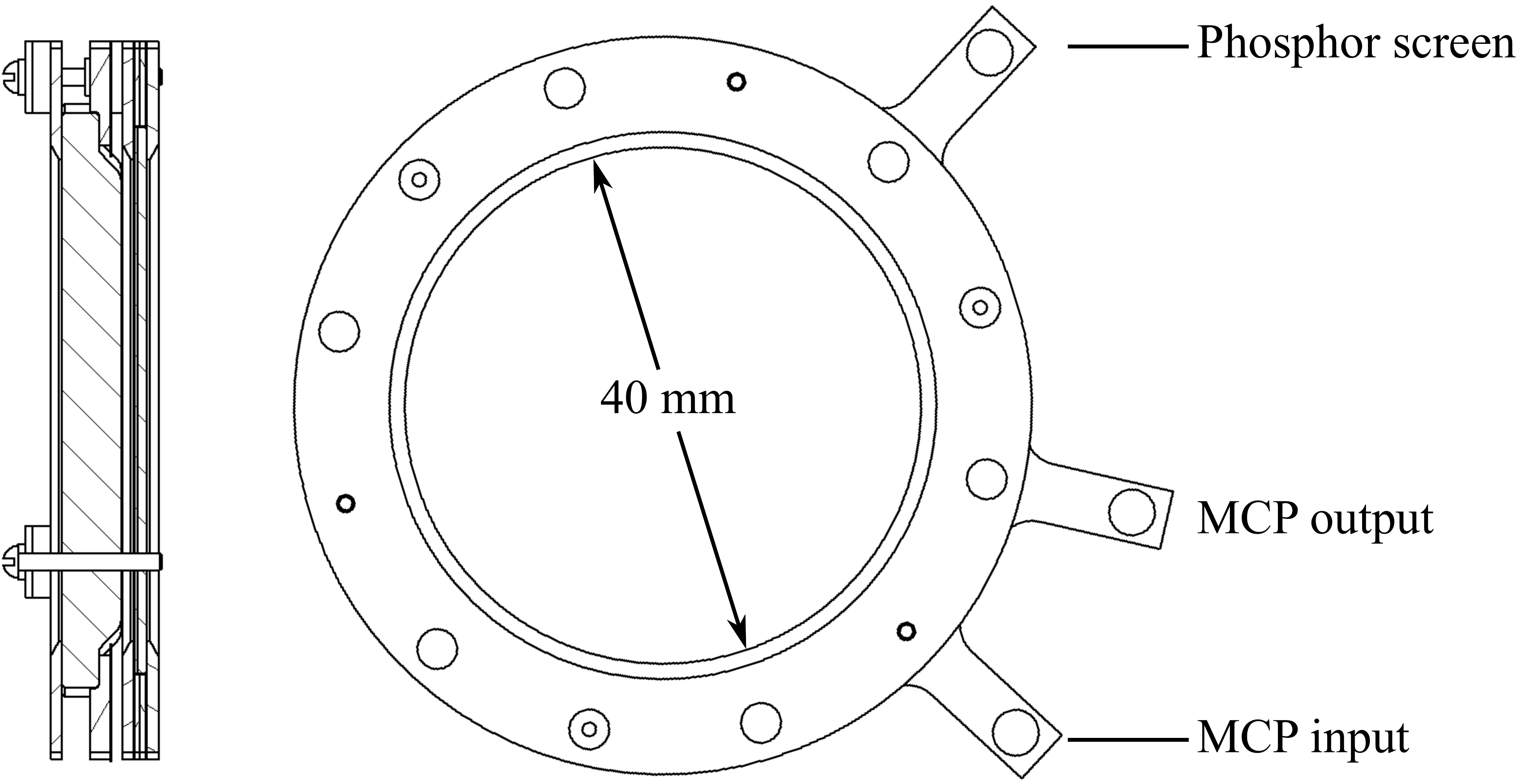}
	\caption{Design drawing of the neutron imaging \textsc{mcp} assembly, consisting of one neutron-sensitive \textsc{mcp}, one regular \textsc{mcp} and a P46 phosphor screen carried by an \textsc{ito}-coated fiber optic faceplate.\label{drawing}}
\end{figure}
The neutron sensitive \textsc{mcp}s we use are made of a glass mixture modified to include enriched $^{10}$B and natural Gd, both with a high neutron capture cross-section.
Neutron capture triggers a nuclear reaction: either $\textrm{n}(^{10}\textrm{B}, \alpha)^7\textrm{Li}+\gamma$ (fission) or $\textrm{n}(^{157}\textrm{Gd},^{158}\textrm{Gd})\textrm{e}^-+\gamma$ (internal conversion).
The fission fragments or conversion electrons have a range of a few microns in this glass.
With an 8 \textmu m pore diameter and a 10 \textmu m pore spacing, this range is sufficient to reach the nearest pore, where electrons are emitted in the vacuum.
This starts a cascade of secondary emissions that is the normal working principle of \textsc{mcp}s.
While most detectors made with these \textsc{mcp}s are read out electronically, with high granularity strip or pixel anodes, the detector presented here is coupled to a phosphor screen and read out optically by a camera.
Figure \ref{drawing} shows a drawing of the assembly.

The imager used for these tests used a neutron sensitive \textsc{mcp} followed by a regular \textsc{mcp}; this configuration allows a high gain, but widens the point spread function, thus affecting the imaging resolution.
A fiber optic faceplate coated with an indium tin oxide (\textsc{ito}, a transparent conductor) electrode serves as the substrate on which the phosphor screen is deposited.
The phosphor used is P46 (Y$_3$Al$_5$O$_{12}$:Ce), a fast phosphor emitting in the green.
 \begin{figure}[tb]
	\centering\includegraphics[width=.65\textwidth]{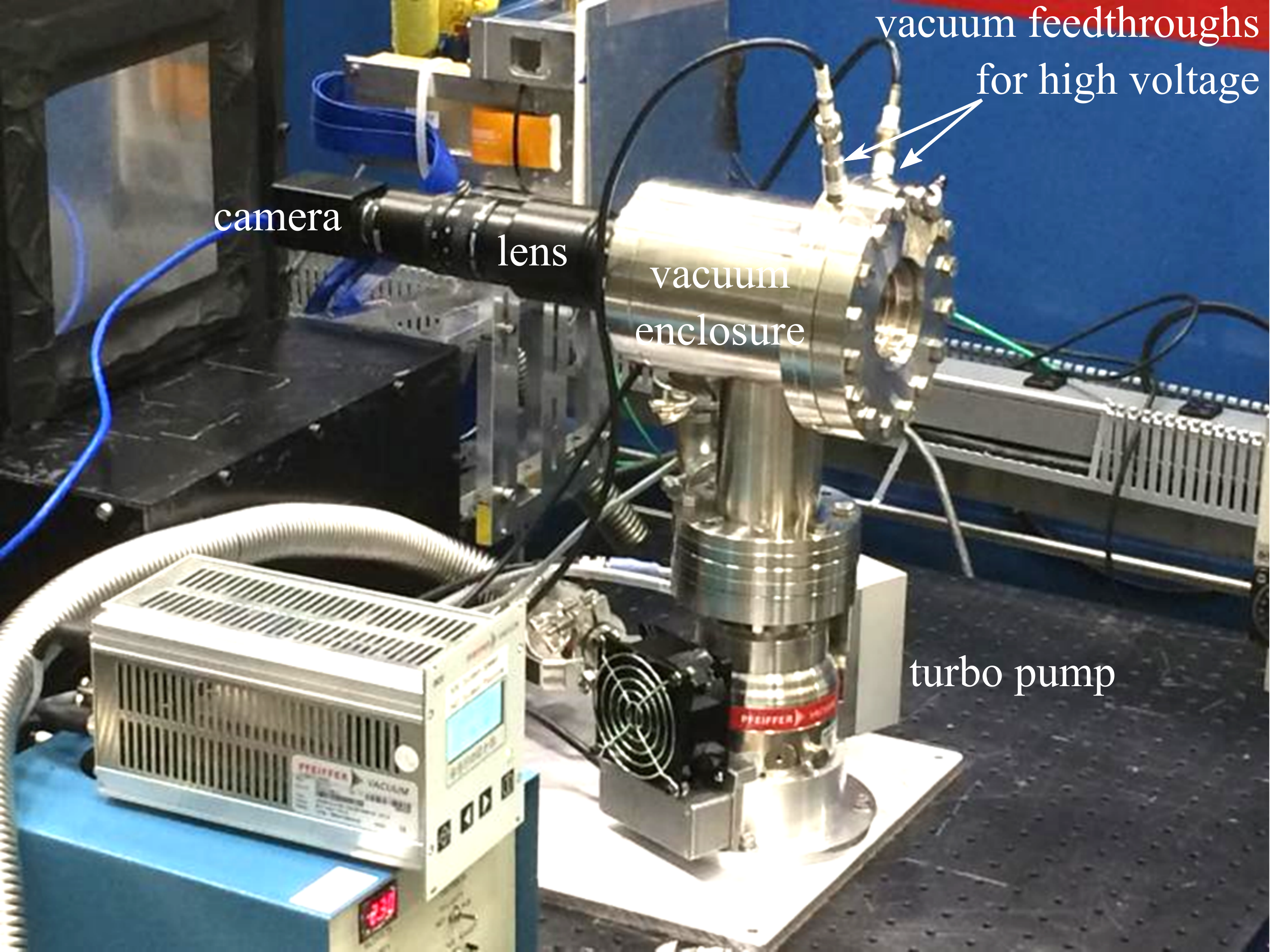}
	\caption{The neutron imager in the CG-1D beam line of \textsc{hfir} at Oak Ridge National Lab. On the bottom left the controller unit of the turbo pump and the high voltage power supply are visible.\label{Delft}}
\end{figure}
The \textsc{mcp}/phosphor screen assembly is installed in a stainless steel vacuum enclosure, a few millimeters from a sapphire viewport which serves as the neutron entrance window.
The remaining volume in the vacuum system is empty, serving only to optimize conductance to the turbo pump just underneath.
The camera\footnote{A Nocturn \textsc{gp} monochrome \textsc{cmos} camera from \textsc{Photonis} was used.} with its relay lens are mounted on another viewport at the back of the vacuum chamber.
This is not ideal, since the camera operates on the neutron beam axis, and is exposed to any neutron flux not absorbed by the \textsc{mcp}; we foresee a mirror mounted at $45^\circ$ to the beam axis in further studies, so the camera and lens will be protected from the neutron flux.

\subsection{Acquiring neutron images}
The high gain of the double \textsc{mcp}, combined with the good sensitivity and limited well depth of the \textsc{cmos} sensor (25000 e$^-$), causes pixel saturation at long exposure.
\begin{figure}[tb]
	\centering\includegraphics[width=\textwidth]{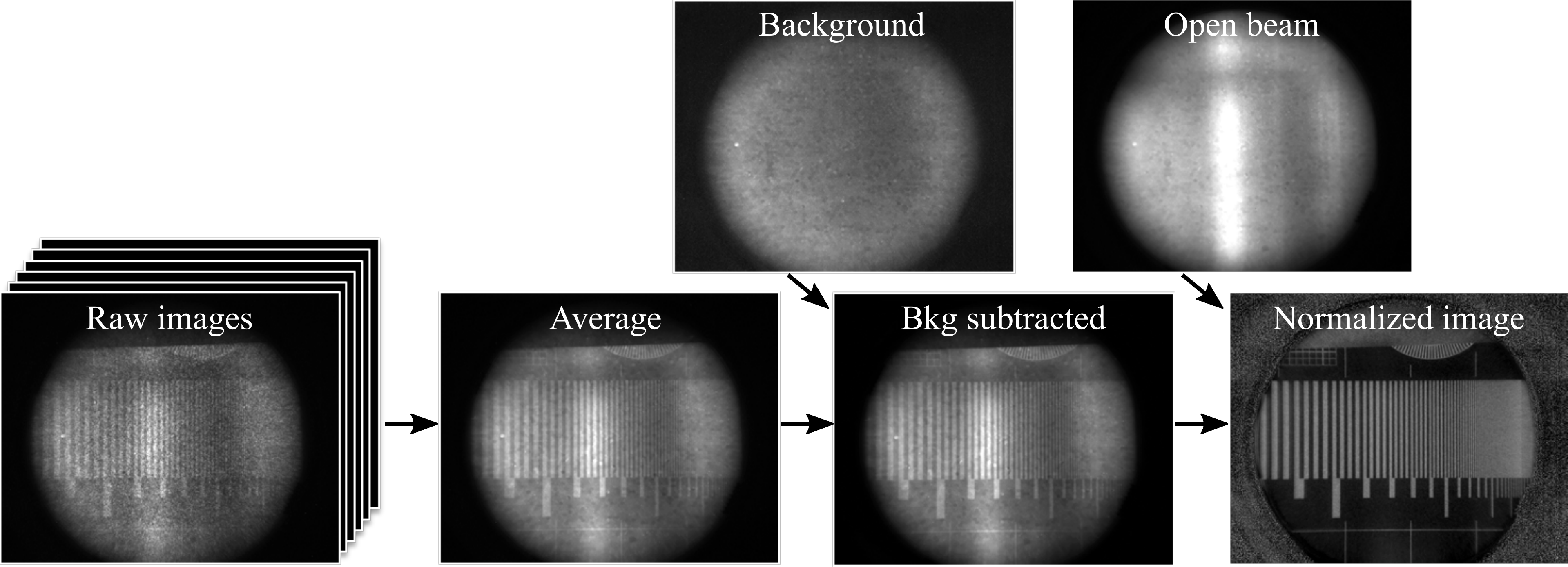}
	\caption{The processing done to generate normalized neutron images.\label{processing}}
\end{figure}
Therefore several frames of short exposure ($<1$ s) are acquired, which are then averaged to obtain a longer exposure image.
The gray value of each pixel of the image must be normalized to a full scale running from the background level (black) to the value obtained with the neutron beam on, but the sample removed (white):
\begin{equation}
I_\text{norm}=\frac{I_\text{raw}-I_\textsc{bkg}}{I_\textsc{ob}-I_\textsc{bkg}},
\end{equation}
with the suffixes \textsc{bkg} and \textsc{ob} meaning background and open beam, respectively.
This cancels the effects of any non-uniformity of backgrounds, the beam profile and the gain of the \textsc{mcp}s.
Figure~\ref{processing} shows this process and its effect on the image.
These operations and all other image processing in this work were done with the open source ImageJ~\cite{ImageJ} software.

\section{Results}
We will discuss a series of images taken primarily with the aim to reveal some of the performance characteristics of this prototype.

\subsection{Gadolinium test mask}
\begin{figure}[b]
	\centering\includegraphics[width=.9\textwidth]{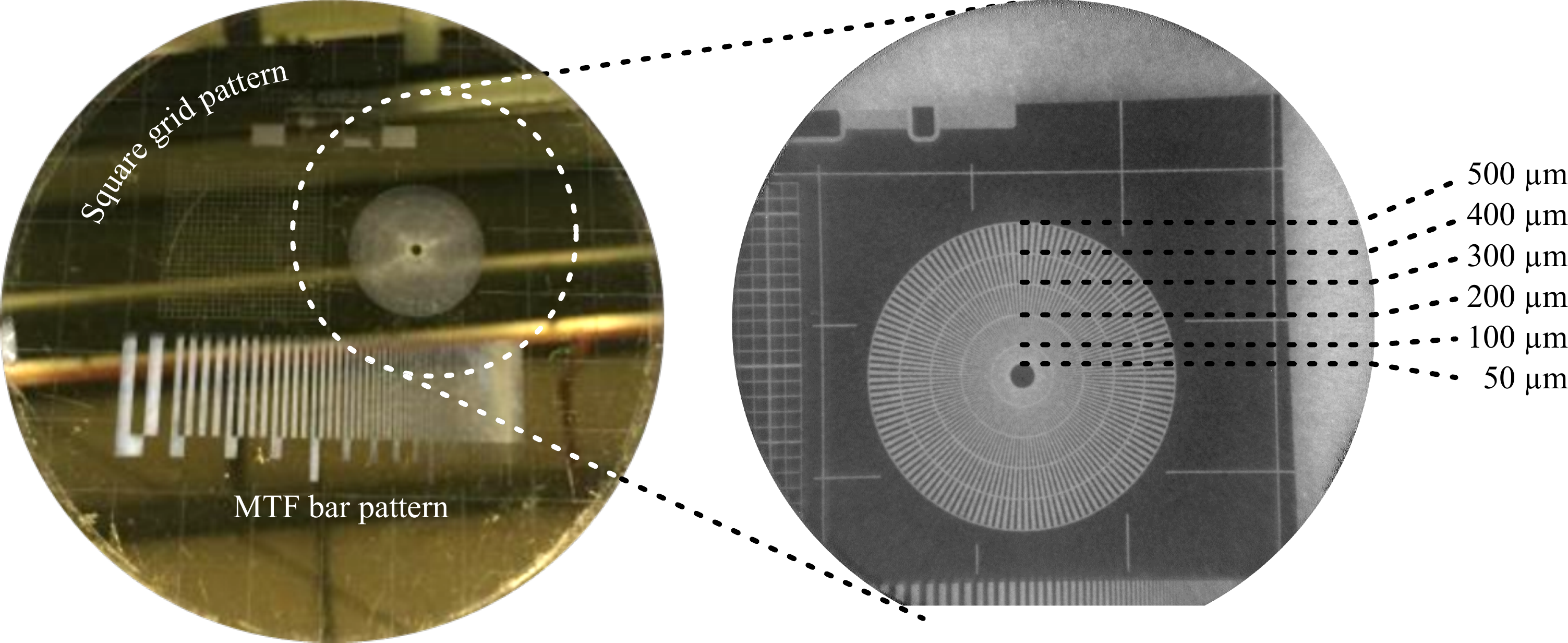}
	\caption{Neutron image of a Siemens star on the Gd mask.\label{Gd-mask}}
\end{figure}

The neutron imaging group at the Paul Scherrer Institute in Switzerland has developed high definition gadolinium patterns using a silica wafer as a substrate~\cite{Gd-mask}, see figure~\ref{Gd-mask}, left.
These test masks are helpful tools to quantify specific performance aspects of a detector.
The images from this mask discussed below were taken at the CG-1D cold neutron imaging beamline of \textsc{hfir} at Oak Ridge National Lab.

\paragraph{Siemens star}
Figure~\ref{Gd-mask} shows a photograph of a wafer with the Gd pattern (left).
There are 3 relevant patterns on the 100 mm diameter wafer, the neutron image on the right shows the Siemens star.
A few circles mark line pitches as indicated in the figure; with this pattern one can immediately estimate the limiting resolution to be between 200 and 100 \textmu m.
Since the pattern has radial lines in all directions, it will also easily reveal differences in horizontal and vertical resolution.
This could happen due to optical misalignment of the camera and lens behind the phosphor screen, or trivial details such as pump vibrations coupling to the imager.
No such difference is seen here.

\begin{figure}[tb]
	\centering\includegraphics[width=\textwidth]{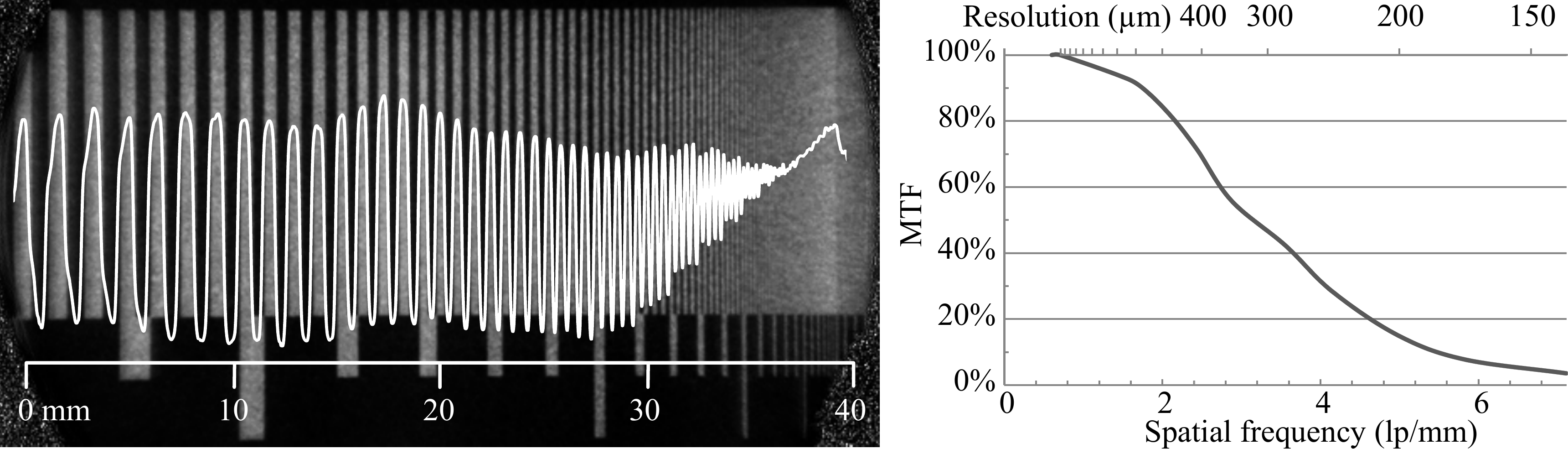}
	\caption{Left: neutron image of the bar pattern to measure the \textsc{mtf}. Overlaid is an intensity profile of the image. Right: \textsc{mtf} curve derived from the intensity profile. \label{BarPattern}}
\end{figure}
\paragraph{\textsc{Mtf} bar pattern}
The pattern of progressively more closely spaced bars (also shown in figure~\ref{processing}) provides a more quantitative way to characterize the detector's spatial resolution, see figure~\ref{BarPattern}.
As the spatial frequency of the bar pattern increases from left to right, the contrast between light and dark shapes diminishes, as clearly shown by the overlaid intensity profile.
The relative contrast as function of the spatial density is called the \emph{Modulation Transfer Function} (\textsc{mtf}), see the curve on the right.
The \textsc{mtf} is equal to the Fourier transform of the line spread function, but the bar pattern provides a convenient way of measuring it.
The \emph{limiting resolution} is the finest spacing that can still be resolved; although somewhat arbitrary, one often equates it to the resolution at 5\% \textsc{mtf}.
Using this standard, the limiting resolution of this detector is just below 150 \textmu m.
It is known from other \textsc{mcp}/phosphor screen detectors (e.g. image intensifier tubes) that \textsc{mcp}s have a non-Gaussian line spread function, which causes the \textsc{mtf} to fall rather gradually; scintillator imagers on the other hand have an \textsc{mtf} that drops quite steeply.

\begin{figure}[b]
	\centering\includegraphics[width=.8\textwidth]{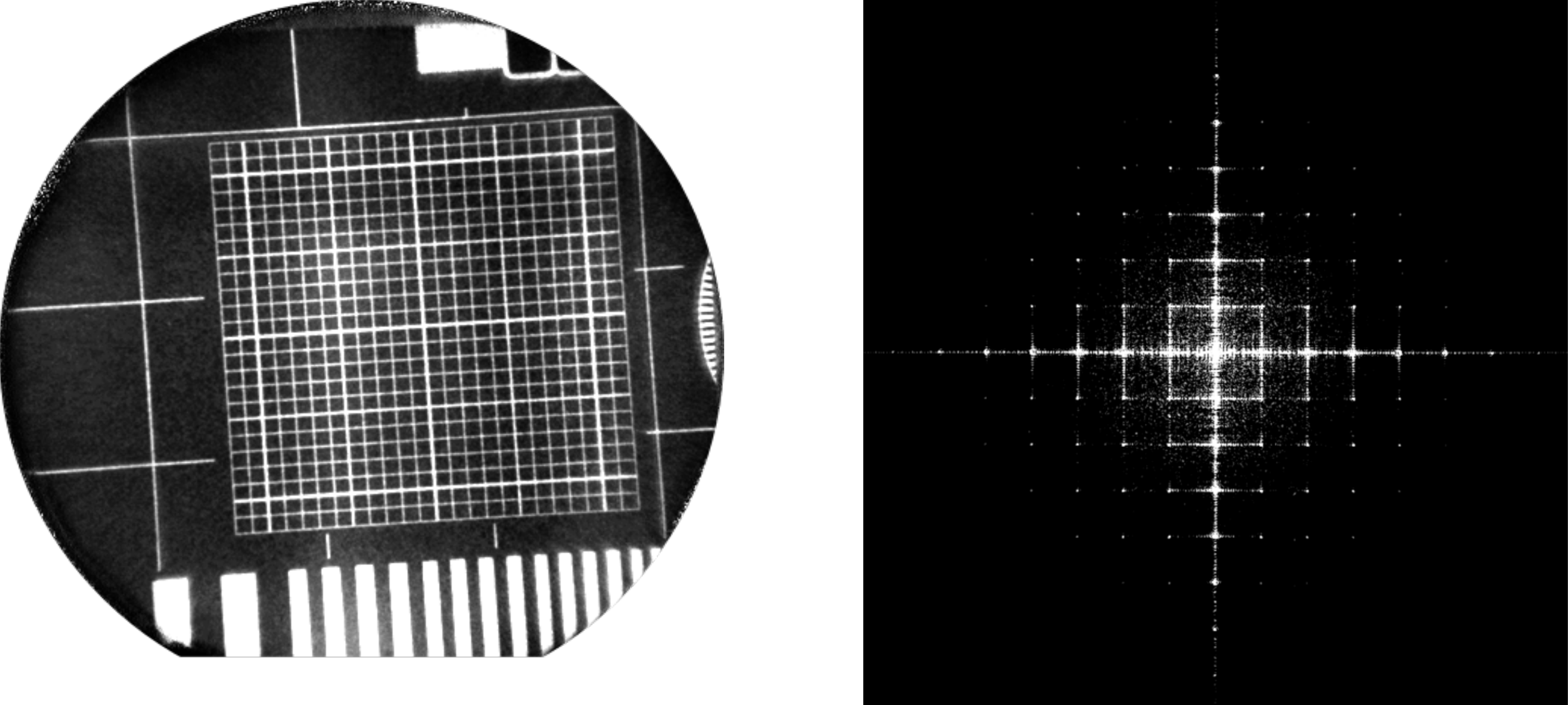}
	\caption{The square grid pattern (left) can reveal image distortions. A discrete Fourier transform of this image (right) is very sensitive to any such distortions.\label{SquareGrid}}
\end{figure}
\begin{figure}[tb]
	\centering\includegraphics[width=.65\textwidth]{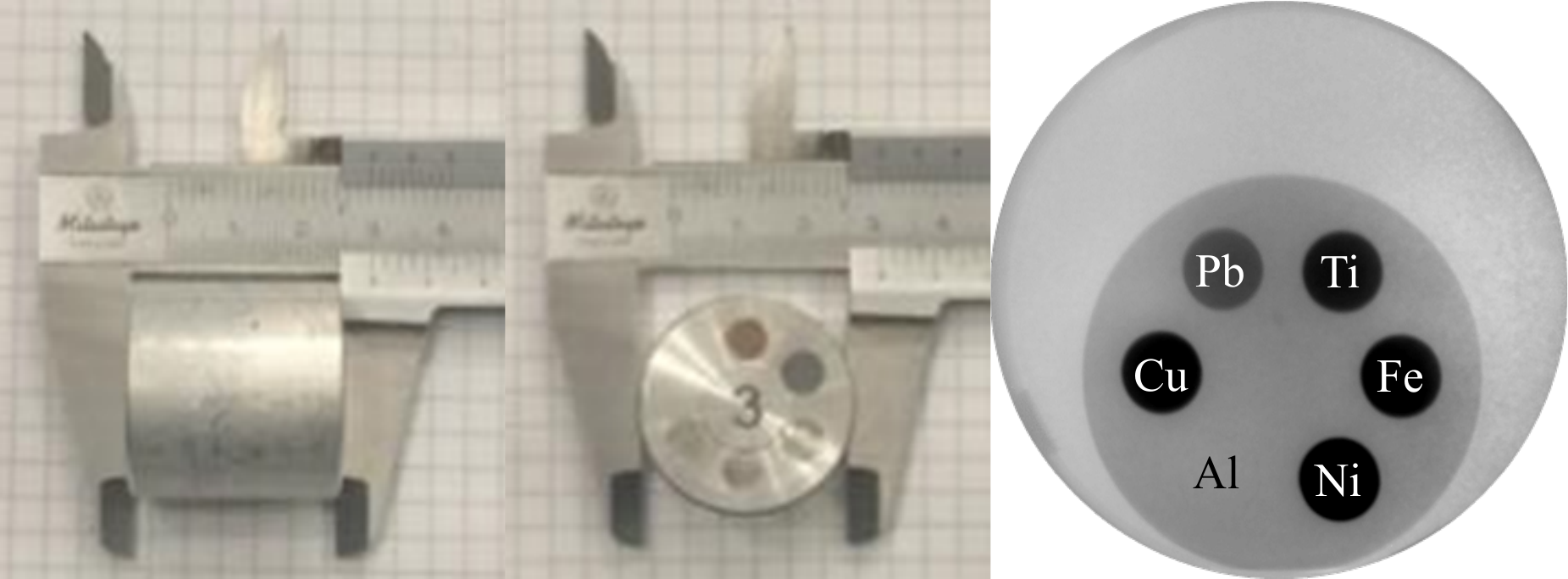}
	\caption{Left, an aluminium cylinder with inserts of several other metals. Right, a neutron image of this cylinder.\label{cylinder}}
\end{figure}
\paragraph{Square grid pattern}
This pattern is a checked square with an even density of lines with varying thickness, see figure~\ref{SquareGrid}, left.
The line spacing is 1 mm, and the lines are 50, 100 and 150 \textmu m wide.
This pattern provides a means of measuring distortions, such as pincushion or barrel distortions.
Any such distortions will normally be due to the optics behind the phosphor screen and not caused by the imager itself.
A 2D discrete Fourier transform of the image will reveal any distortions clearly, as the pattern of spots will smear out if there is any distortion.
To the right of figure~\ref{SquareGrid} is a 2D Fourier transform, showing no distortions.
The fact that the active area of 40 mm diameter is rather small compared to many scintillator screens may help here, as well as the absence of a mirror to reflect the image away from the neutron beam axis.

\subsection{Neutron tomography}
Computed tomography is a technique where the benefit of much greater sensitivity compared to scintillator screens is particularly helpful.
A sample is mounted on a rotation stage and images are taken at many hundreds of different angles.
With the sensitivity of scintillating screens, the acquisition of so many images takes many hours, and often needs to be done overnight.
With this \textsc{mcp}-based neutron imager we can do a scan of similar quality in two hours or less.
Figure~\ref{cylinder} shows a test device developed at PSI~\cite{tomography} to evaluate the contrast revealed by a neutron imaging beamline and its detector.
It is an aluminum cylinder, with smaller cylindrical inserts of other metals.
To the right is a neutron image taken at the imaging beamline of the Reactor Institute of Delft University of Technology; the elements of the inserts are indicated.
\begin{figure}[tb]
	\centering\includegraphics[width=\textwidth]{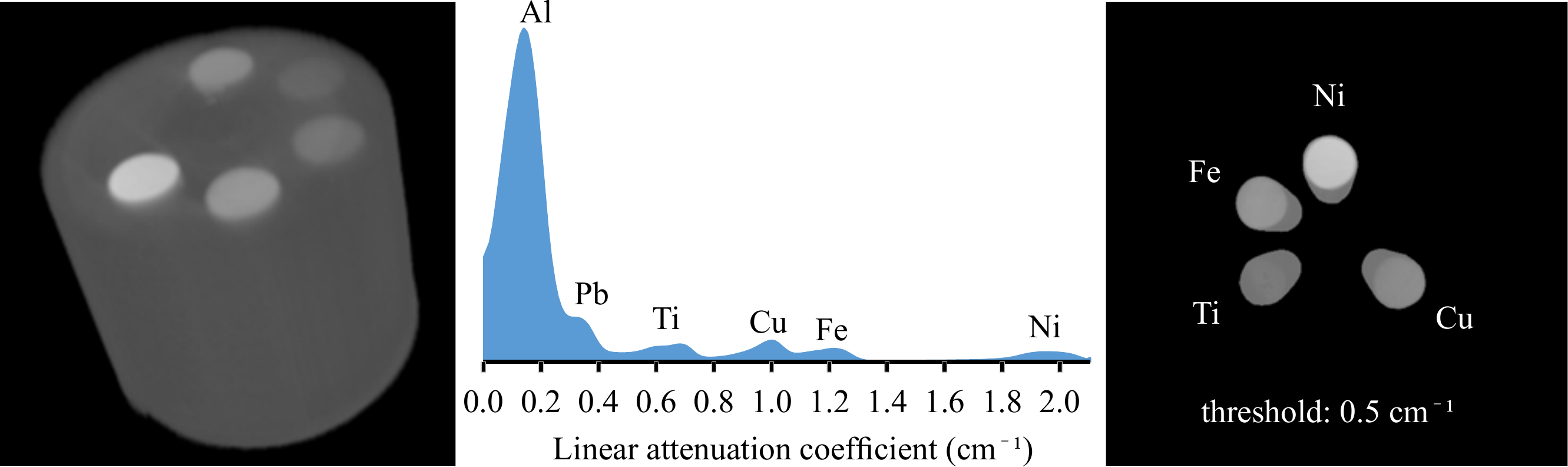}
	\caption{Left, tomographic reconstruction of the test object from figure~\ref{cylinder}. Right, histogram of greyscales of the 3D voxel map. The horizontal axis expresses the grey values as attenuation coeffient.\label{tomography}}
\end{figure}

Figure~\ref{tomography} shows a tomographic reconstruction of the same object.
The same algorithms used for X-ray CT are employed to reconstruct a voxel map of attenuation coefficients\footnote{the shown reconstruction was done with Octopus 8.8.1.0, a software package designed for X-ray CT.} from the many images, 900 projections in this case.
In the center of figure~\ref{tomography} is a histogram of the grey values in the 3D voxel map of the recontruction, set to a scale of attenuation coefficients.
The six different materials present in the sample are clearly visible as six peaks of equal area, except for the aluminum peak from the cylinder itself.
The attenuation coefficients of the six peaks are consistent with literature values for these metals.
Remarkable is the low attenuation of lead, which has great stopping power for X-rays; for thermal neutrons it is hardly more absorbing than aluminum.
This histogram also provides the key to "seeing through" samples, one of the powerful tools tomography offers.
If in the 3D reconstruction image one puts limits on what is displayed or not, certain materials can be rendered invisible so as to only show the materials of interest.
The right of figure~\ref{tomography} shows the same reconstruction where only voxels over a threshold of 0.5 cm$^{-1}$ are displayed.
From the histogram it is clear that this should exclude the aluminum cylinder and the lead insert, and show the other four inserts.

\section{Conclusions and outlook}
\begin{figure}[b]
	\centering\includegraphics[width=\textwidth]{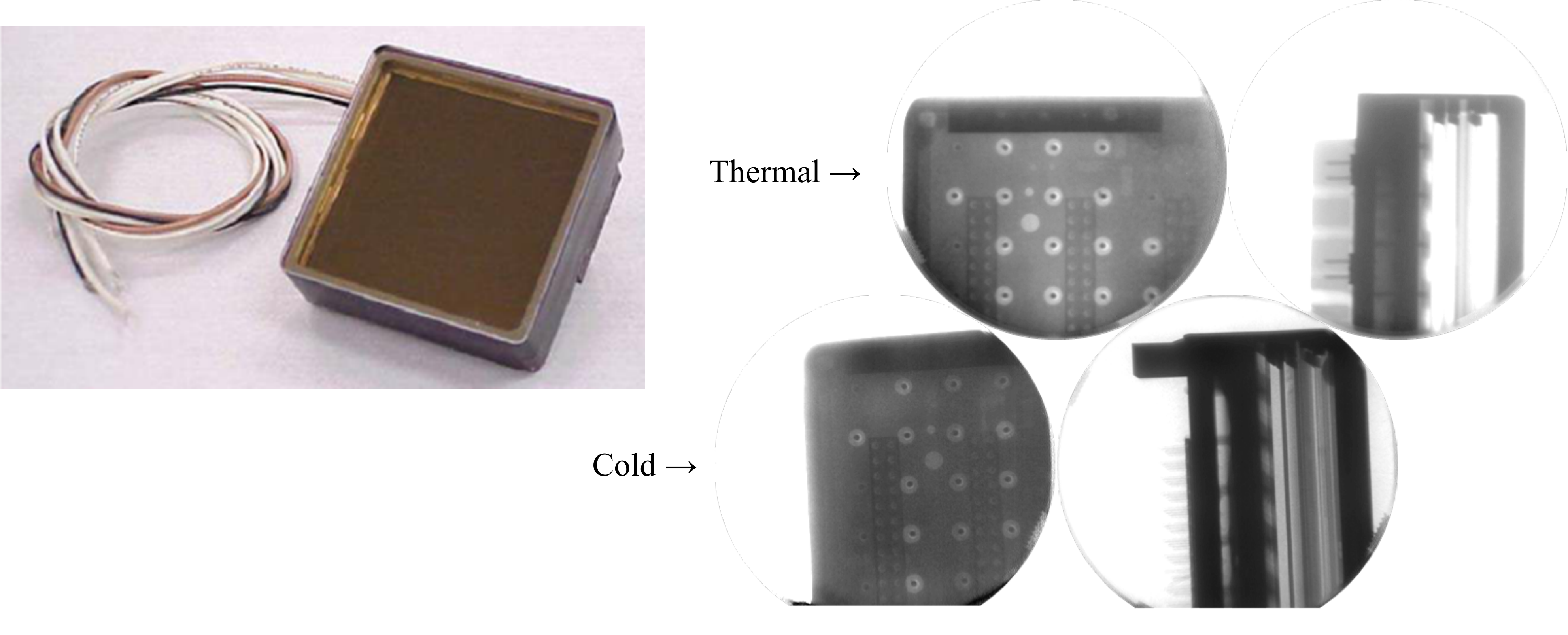}
	\caption{Neutron images of a Planacon$^\textsc{tm}$ \textsc{mcp-pmt} with thermal neutrons (Delft) and with cold neutrons (Oak Ridge). This highlights the difference between thermal and cold neutron imaging: thermal neutrons have deeper penetration, but cold neutrons give more contrast.\label{thermal_vs_cold}}
\end{figure}
Tests done at two different beamlines with different neutron energies (see figure~\ref{thermal_vs_cold}) have shown that this first prototype performs well for neutron imaging and tomography.
The sensitivity is orders of magnitude higher than conventional scintillator-based imaging detectors.
The resolution, although sufficient for most imaging applications, is not yet as good as the best scintillator detectors.
There is also room for improvement on practical aspects.
After being tested extensively, the stainless steel vacuum system of the imager shown here is strongly activated, contributing significantly to its own gamma background.
The setup consists of many parts, pumps, controllers, power supplies etc., all interconnected by a web of cables and hoses.

An improved \textsc{mcp} based imager, already completed at the time of writing, has a square active area of $100\times100$ mm$^2$, and attains a limiting resolution of better than 50 \textmu m.
A square-shaped field of view is advantageous when doing tomography, resulting in a cylindrical fiducial volume, rather than the spherical fiducial volume of the round imager in this work.
Its choice of materials and construction has eliminated the activation issues, and all loose parts are located in one control unit, with a \textsc{usb} connection to a computer to operate it remotely.

A neutron imaging detector with a sensitivity so much higher than present-day techniques may enable new sorts of imaging studies.
Neutron tomography can be done so quickly that many samples can be scanned the same day, making more efficient use of precious beam time.
Neutron imaging can also be applied to dynamic processes, opening the door to neutron video recording.

The fact that many times less neutron flux is integrated to attain a certain image quality also comes with benefits.
Samples activate less, proportionally to exposure time.
Rare artifacts and valuable museum pieces can be imaged and still return to their owner.
Small, low power nuclear reactors running on conventional low-enriched uranium become suitable neutron sources for imaging.
The images in this study taken at the reactor in Delft are a case in point.
We are exploring the possibility of neutron imaging with neutron generators, which may take neutron imaging from large scale user facilities to labs in academia and industry.

\acknowledgments
We would like to acknowledge the Reactor Institute Delft (\textsc{rid}), Delft University of Technology, for the beam time and the support during the measurements.
All the work presented here with thermal neutrons was done at the \textsc{rid}.

All research with cold neutrons was done at the CG-1D beamline of the High Flux Isotope Reactor (\textsc{hfir}), a \textsc{doe} Office of Science User Facility operated by the Oak Ridge National Laboratory.

\end{document}